\documentstyle[twoside,12pt]{article}
\normalsize
\newcommand{\bdi}{\begin{displaymath}}
\newcommand{\edi}{\end{displaymath}}
\newcommand{\bfi}{\begin{figure}}
\newcommand{\efi}{\end{figure}}

\newcommand{\beq}{\begin{equation}}
\newcommand{\eeq}{\end{equation}}
\newcommand{\beqa}{\begin{eqnarray}}
\newcommand{\eeqa}{\end{eqnarray}}
\newcommand{\no}{\nonumber}

\newcommand{\dsla}{\partial\hspace{-6pt} /  }  
\newcommand{\Asla}{A\hspace{-6.5pt}  /  }

\newcommand{\wt}{\widetilde}

\begin{document}
\begin{titlepage}
\begin{flushright}
\today
\end{flushright}

\vspace{1cm}
\begin{center}
{\Large \bf Renormal-order improvement of the Schwinger mass}\\[1cm]
C. Adam* \\
School of Mathematics, Trinity College, Dublin 2 \\

\vfill
{\bf Abstract} \\
\end{center}
The massive Schwinger model may be analysed by a perturbation expansion in
the fermion mass. However, the results of this mass perturbation theory are
sensible only for sufficiently small fermion mass. By performing a 
renormal-ordering, we arrive at a chiral perturbation expansion where the
expansion parameter remains small even for large fermion mass. We use this
renormal-ordered chiral perturbation theory for a computation of the
Schwinger mass and compare our results with lattice computations.

\vfill

$^*)${\footnotesize  
email address: adam@maths.tcd.ie, adam@pap.univie.ac.at}
\end{titlepage}

\section{Introduction}

The Schwinger model \cite{Sc1,LS1}, or two-dimensional QED with one fermion
flavour, is one of the most widely studied quantum
field theoretic models. It has the attractive feature of being rather
simple, while, nevertheless, it has many properties similar to those of
more complicated gauge theories. Among them are the chiral anomaly 
\cite{Ad1}--\cite{Bertl}, the
formation of a chiral condensate and the non-trivial vacuum structure
($\theta$-vacuum) \cite{LS1,AAR}, \cite{HeHo1}--\cite{Adam}. 
In addition, the fundamental fermion is confined, and
only mesonic states occur in the physical particle spectrum \cite{Sc1,LS1}.

The Lagrangian density of the model is
\beq
L= \bar\Psi (i\dsla -e \Asla -m)\Psi -\frac{1}{4}F_{\mu\nu}F^{\mu\nu} .
\eeq
When the fermion is massless ($m=0$), the model may be solved exactly, and
it turns out to be equivalent to the theory of one free, massive boson field
(``Schwinger boson'') with Schwinger mass $\mu =e/\sqrt{\pi}$. 

For the massive ($m\ne 0$) Schwinger model, the Schwinger boson turns into
an interacting particle, and its mass acquires corrections. The investigation
of the massive Schwinger model started more than 20 years ago with a
series of now classical papers \cite{CJS}--\cite{FS1} and has continued
since then (see \cite{AAR,MSMPT} for a review). 
Quite recently, the corrections to the Schwinger mass, as well
as the formation of higher bound states and the chiral condensate in the
massive Schwinger model, have been studied by a variety of methods. Among 
these methods are mass perturbation theory \cite{MSMPT}--\cite{SCAT}, 
light-front quantization \cite{EP1}--\cite{Burk1}, lattice computations
\cite{CH1}--\cite{FHTS}, and a generalized Hartree-Fock approach 
\cite{HoRo1}--\cite{HoRo3}. The multi-flavour case has been discussed, e.g.,
by \cite{HHI1}--\cite{Gatt2}.

In this article we want to compute the Schwinger mass of the massive 
Schwinger model by a method that is similar to mass perturbation theory.
By performing a renormal-ordering, we shall find a new expansion parameter
(instead of the fermion mass $m$) that remains rather small even for large
$m$. Therefore, the resulting chiral perturbation expansion will produce
a Schwinger mass that tends to the result of mass perturbation theory for
small fermion mass $m$, but remains reasonable even for large $m$. 
Finally, we shall compare our results to some lattice data.

We use the bosonized, Euclidean version of the massive Schwinger model in
the sequel, and our conventions for the massive scalar propagator are
\beq
D_\mu (x) =-\frac{1}{2\pi}K_0 (\mu|x|),\qquad \wt D_\mu (p)=
\frac{-1}{p^2 +\mu^2} ,
\eeq
where $K_0$ is the McDonald function.

\section{Renormal-ordered chiral perturbation theory}

Our starting point is the Euclidean, bosonized Lagrangian density of the
massive Schwinger model \cite{CJS,Co1,FS1},
\beq
L=-N_\mu \Bigl[ \frac{1}{2} \Phi (\Box -\mu^2)\Phi +\frac{e^\gamma}{2\pi} \mu 
m\cos (\sqrt{4\pi}\Phi +\theta )\Bigr] .
\eeq
Here $\mu =e/\pi^{1/2}$ is the Schwinger mass of the massless ($m=0$) 
Schwinger model, $\theta$ is the vacuum angle and $\gamma =0.5772$ is the
Euler constant. Further, $N_\mu$ denotes normal-ordering w.r.t. $\mu$.

By making use of the well-known normal-ordering relation 
\beq
N_{\bar\mu} e^{\pm i\beta \Phi (x)}=
(\frac{\mu}{\bar\mu})^{\frac{\beta^2}{4\pi}}
N_\mu e^{\pm i\beta \Phi (x)}
\eeq
(see \cite{Co2}), we may rewrite the Lagrangian density (3) for arbitrary
normal-ordering mass $\bar\mu$ like
\beq
L= -N_{\bar\mu} \Bigl[ \frac{1}{2} \Phi (\Box -\mu^2)\Phi +
\frac{e^\gamma}{2\pi}\bar\mu m\cos (\sqrt{4\pi}\Phi +\theta)\Bigr]
\eeq
up to an irrelevant constant. Next, let us observe that the interaction
term $\sim \cos\sqrt{4\pi}\Phi$ contains a $\Phi^2$ piece, too. We now
want to shift this $\Phi^2$ piece from the interaction term to the mass
term in (5), and we want to choose the normal-ordering mass $\bar\mu$ 
equal to the total mass of the resulting mass term. We find the equation
\beq
\bar\mu^2 = \mu^2 +2e^\gamma \cos\theta \, \bar\mu m
\eeq
with the solution
\beq
\bar\mu = e^\gamma \cos\theta \, m + \sqrt{\mu^2 + e^{2\gamma} \cos^2 \theta
\, m^2} .
\eeq
Our new Lagrangian density reads
\beqa
L&=& -N_{\bar\mu} \Bigl[ \frac{1}{2}\Phi (\Box -\bar\mu^2)\Phi +\lambda
(\frac{1}{2}(e^{i\theta}e^{i\sqrt{4\pi}\Phi} + e^{-i\theta}e^{-i\sqrt{4\pi}
\Phi}) + 2\pi\cos\theta \, \Phi^2 -\cos\theta)\Bigr] \no \\
&=:& L_0 +L_{\rm I}
\eeqa
\beq
\lambda =\frac{e^\gamma}{2\pi}m\bar\mu ,
\eeq
where we have subtracted a constant ($\cos\theta$) in the interaction term 
in order to avoid infinite vacuum energy contributions to the perturbation
series. Further, we rewrote the sine and cosine of the field $\Phi$ in
terms of exponentials, because the latter ones have simpler Wick contractions
(see (12) below). By this simple rewriting of the Lagrangian density
we have achieved 
a major advantage. A perturbative expansion of the original version (3) of
the Lagrangian is an expansion in powers of $m$ and will, obviously, only
lead to sensible results for sufficiently small $m$. On the other hand,
for the new version (8) of the Lagrangian, the (dimensionless) expansion
parameter of the perturbation expansion is 
\beq
\frac{\lambda}{\bar\mu^2} = \frac{e^\gamma m}{2\pi\bar\mu}=
\frac{e^\gamma m}{2\pi (e^\gamma \cos\theta \, m+ \sqrt{\mu^2 +e^{2\gamma}
\cos^2 \theta \, m^2})} ,
\eeq
which is rather small for all $m$ provided that the vacuum angle $\theta$ is
not too large. E.g., for $\theta=0$ we find
\beq
0=(\frac{\lambda}{\bar\mu^2})(m=0,\theta=0)<(\frac{\lambda}{\bar\mu^2})
(m,\theta =0)<(\frac{\lambda}{\bar\mu^2})(m=\infty ,\theta =0)=\frac{1}{4\pi}
\eeq
(see Fig. 1). 

\input psbox.tex 
\begin{figure}
$$\psbox{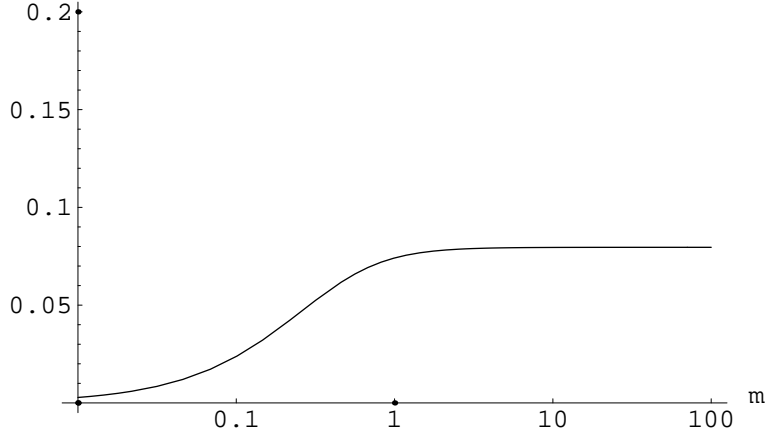}$$
\caption{The figure shows how the expansion parameter 
$\frac{\lambda}{\bar\mu^2}$ ($y$-axis) varies with the fermion mass $m$.
We use units $\mu =1$.} 
\end{figure}

Therefore, the perturbative expansion for the theory defined
by (8) leads to finite results for all $m$ and, as the expansion parameter 
is quite small, one may hope that these results are at least qualitatively 
correct even for large $m$. The interaction Lagrangian in (8) contains a
$\Phi^2$ piece in addition to the exponentials, therefore the resulting
chiral perturbation theory is slightly more complicated than the usual
mass perturbation expansion, because it contains Wick contractions among
both polynomials and exponentials in $\Phi$. All these may be inferred from
the general formula   (see \cite{MSMPT,Gatt1,Gatt2})
\bdi
\langle \prod_{l=1}^n e^{i\sigma_l \beta \Phi (x_l)}\rangle_0 [\rho]
=e^{\sum_{i<j}\sigma_i \sigma_j \beta^2 D_{\bar\mu}(x_i -x_j)} \cdot
\edi
\beq
\cdot e^{-\frac{1}{2}\int dy_1 dy_2 \rho (y_1) D_{\bar\mu}(y_1 -y_2)\rho (y_2)
+i\beta \sum_{k=1}^n \sigma_k \int dy \rho (y) D_{\bar\mu}(y-x_k)} ,
\eeq
where $\sigma_l =\pm 1$ and $\rho (x)$ is a source term for the $\Phi (x)$
field (i.e., $\Phi (x_i)$ are inserted into the VEV (12) by taking functional
derivatives $\delta /\delta \rho (x_i)$). The subscript 0 means $\lambda =0$,
and $\beta =\sqrt{4\pi}$ in our case. Further, the exponentials on the l.h.s.
are normal-ordered w.r.t. the mass of the $\Phi$ field, $\bar\mu$. 

The Lagrangian density (8), together with the Wick contraction formula (12), 
will now serve as a starting point for the perturbative computation of the
two-point function $\langle \Phi (y_1) \Phi (y_2)\rangle_\lambda $ and for
the approximate determination of the Schwinger mass.

\section{Two-point function and Schwinger mass}

Before actually starting the computation, let us observe that already the
zeroth order Schwinger mass $\bar\mu$ gives a certain prediction for large
$m$ if $\theta =0$ (numerical estimates for large $m$ will always be done 
for $\theta =0$, because for $\theta \ne 0$ the expansion parameter
$\lambda / \bar\mu^2$ is larger in the large $m$ region, making thereby 
numerical estimates less reliable). We find
\beq
\lim_{m\to\infty} \bar\mu (m,\theta =0) = 2e^\gamma m \simeq 3.56 m .
\eeq
The true value is $2m$, because in the limit of vanishing (electromagnetic)
interaction the Schwinger boson $\Phi$ is just composed of two free fermions,
each of mass $m$. Therefore, this lowest order estimate is nearly 80\% off
the exact value (which is not too bad for a lowest order estimate). 
Further, when Taylor-expanded in $m$, $\bar\mu$ reproduces the first order
correction of mass perturbation theory ($M_0$ $\ldots$ Schwinger mass in zeroth
order)
\beq
M_0 := \bar\mu = \mu + e^\gamma \cos\theta \, m + o(\frac{m^2}{\mu^2}) .
\eeq
This feature should generalize to higher orders: The two perturbation 
expansions (in $m$ for the Lagrangian (3); in $\lambda$ for the Lagrangian (8))
are just different formulations of the same theory; further, the parameter 
$\lambda$ (or equivalently $\lambda /\bar\mu^2$) has a well-defined Taylor
expansion in $m$, starting with a linear ($m^1$) term. Therefore, the order
$n$ result of the perturbation series in $\lambda$, when expanded in $m$ 
up to $n$-th order, should coincide with the $n$-th order mass perturbation
theory result. We conclude that there is no $\lambda^1$ contribution to
the $\lambda$ perturbation series for the Schwinger mass.

The perturbation expansion for the two-point function is
\beq
\langle \Phi (y_1)\Phi (y_2)\rangle_\lambda =\frac{1}{Z}\langle \Phi (y_1)
\Phi (y_2) \sum_{n=0}^\infty \frac{\lambda^n}{n!} \prod_{l=1}^n \int dx_l
L_{\rm I} (x_l)\rangle_0 ,
\eeq
but we prefer not to include the $1/Z$ normalization factor explicitly, and
we will remove disconnected pieces by hand. Therefore we want to compute
\beq
 \langle \Phi (y_1)\Phi (y_2)\rangle_\lambda =\langle \Phi (y_1)\Phi (y_2)
\rangle_0 +
\langle \Phi (y_1)
\Phi (y_2) \sum_{n=1}^\infty \frac{\lambda^n}{n!} \prod_{l=1}^n \int dx_l
L_{\rm I} (x_l)\rangle_0 .
\eeq
First we observe that the $\lambda^1$ contribution is indeed zero. This 
may be checked by explicit computation, but it can be understood immediately
from the following observation. When only one $L_{\rm I}$ is inserted into
the two-point function (16), then for the exponentials $\exp \pm i\sqrt{4\pi}
\Phi (x)$ in $L_{\rm I}(x)$ only the $\Phi^2$ part is effective, because a
higher number of boson lines emerging from the vertex $L_{\rm I}$ cannot
be Wick-contracted to anything. But precisely this $\Phi^2$ piece is subtracted
in $L_{\rm I}$, (8), therefore the net contribution is zero.

Up to second order, we obtain
\bdi
\langle \Phi (y_1)\Phi (y_2)\rangle_\lambda = -D_{\bar\mu} (y_1 -y_2 ) 
+\frac{\lambda^2}{8} \int dx_1 dx_2 \langle \Phi (y_1 )\Phi (y_2)\cdot
\edi
\bdi
\cdot (e^{i\theta}e^{i\sqrt{4\pi}\Phi (x_1)}+e^{-i\theta}e^{-i\sqrt{4\pi}
\Phi (x_1)} +4\pi\cos\theta \, \Phi^2 (x_1) -2\cos\theta)\cdot
\edi
\beq 
\cdot (e^{i\theta}e^{i\sqrt{4\pi}\Phi (x_2)}+e^{-i\theta}e^{-i\sqrt{4\pi}
\Phi (x_2)} +4\pi\cos\theta \, \Phi^2 (x_2) -2\cos\theta) ,
\eeq
where we now have to perform all the Wick contractions with the help of
formula (12). Using the symmetry $D_{\bar\mu} (x)=D_{\bar\mu} (-x)$ and the
relabelling symmetry $\int dx_1 dx_2 f(x_1 , x_2)=\int dx_1 dx_2 f(x_2 ,x_1)$
and omitting the disconnected pieces, we arrive at (c \ldots  connected)
\bdi
\langle \Phi (y_1)\Phi (y_2)\rangle_\lambda^{\rm c} =-D_{\bar\mu}(y_1 -y_2)
+2\pi\lambda^2 \int dx_1 dx_2 \cdot
\edi
\bdi
\cdot \Bigl[ -\cos 2\theta e^{4\pi D_{\bar\mu}(x_1 -x_2)}(D_{\bar\mu}(y_1 -
x_1)D_{\bar\mu}(y_2 -x_2) +D_{\bar\mu}(y_1 -x_1)D_{\bar\mu}(y_2 -x_1))
\edi
\bdi
-e^{-4\pi D_{\bar\mu}(x_1 -x_2)}(-D_{\bar\mu}(y_1 -x_1)D_{\bar\mu}(y_2 -x_2)
+D_{\bar\mu}(y_1 -x_1)D_{\bar\mu}(y_2 -x_1))
\edi
\bdi
+(\cos 2\theta +1)(D_{\bar\mu}(y_1-x_1)D_{\bar\mu}(y_2 -x_1) +2\pi 
D_{\bar\mu}(y_1 -y_2)D^2_{\bar\mu}(x_1 -x_2)
\edi
\bdi
+ 8\pi D_{\bar\mu}(x_1 -x_2)D_{\bar\mu}(y_1 -x_1)D_{\bar\mu}(y_2 -x_2) +
8\pi^2 D^2_{\bar\mu}(x_1 -x_2)D_{\bar\mu}(y_1 -x_1)D_{\bar\mu}(y_2 -x_1))
\edi
\bdi
+(\cos 2\theta +1)D_{\bar\mu}(y_1 -x_1)D_{\bar\mu}(y_2 -x_1)
\edi
\bdi
-4\pi (\cos 2\theta +1)D_{\bar\mu}(x_1 -x_2)D_{\bar\mu}(y_1 -x_1)
D_{\bar\mu}(y_2 -x_2)
\edi
\beq
-(\cos 2\theta +1)D_{\bar\mu}(y_1 -x_1)D_{\bar\mu}(y_2 -x_1) \Bigr] ,
\eeq
where in the $\lambda^2$ part the first two terms (with the $\exp \pm
4\pi D_{\bar\mu}(x_1 -x_2)$ factors) are from the $\langle \Phi \Phi
\exp\Phi\exp\Phi\rangle$ contractions, the next term is from the $\langle
\Phi\Phi \Phi^2 \exp\Phi \rangle$ contractions, and the three last terms are
from the $\langle \Phi\Phi\exp\Phi \rangle$, $\langle \Phi\Phi \Phi^2
\Phi^2 \rangle$ and $\langle \Phi\Phi \Phi^2 \rangle$ contractions, 
respectively. It is important now that in expression (18) the exponentials 
$\exp \pm 4\pi D_{\bar\mu}(x_1 -x_2)=1\pm 4\pi D_{\bar\mu}(x_1 -x_2) +\ldots$ 
that are integrated separately
get their constant piece 1 cancelled by some other contributions,
and those exponentials that are convoluted with propagators get their first
order piece $\pm 4\pi D_{\bar\mu}(x_1 -x_2)$ cancelled (their constant
piece may be omitted by hand, because it is a disconnected term). 
Therefore, we arrive at
\bdi
\langle \Phi (y_1)\Phi (y_2)\rangle_\lambda^{\rm c} =-D_{\bar\mu} (y_1 -y_2)
-2\pi \lambda^2 \cdot 
\edi
\bdi
\cdot \Bigl[ \cos 2\theta (E_+ (\bar\mu) \int dx D_{\bar\mu}(y_1 -x)
D_{\bar\mu}(y_2 -x)+ \int dx_1 dx_2 E^{(1)}_+ (x_1 -x_2 ,\bar\mu)
D_{\bar\mu}(y_1 -x_1)D_{\bar\mu}(y_2 -x_2))
\edi
\beq
+E_- (\bar\mu)\int dx D_{\bar\mu}(y_1 -x)
D_{\bar\mu}(y_2 -x)- \int dx_1 dx_2 E^{(1)}_- (x_1 -x_2 ,\bar\mu)
D_{\bar\mu}(y_1 -x_1)D_{\bar\mu}(y_2 -x_2))
\eeq
\beq
-8\pi^2 (\cos 2\theta +1)\int dx_1 dx_2 D^2_{\bar\mu}(x_1 -x_2)D_{\bar\mu}
(y_1 -x_1) D_{\bar\mu}(y_2 -x_1) \Bigr] ,
\eeq
where $E_\pm (\bar\mu), E^{(1)}_\pm (x,\bar\mu)$ are defined like
\bdi
E_\pm (x,\bar\mu) = e^{\pm 4\pi D_{\bar\mu}(x)} -1 , \quad
E_\pm (\bar\mu) =\int d^2 x E_\pm (x,\bar\mu)
\edi
\beq
E^{(1)}_\pm (x,\bar\mu) =e^{\pm 4\pi D_{\bar\mu}(x)}-1 \mp 4\pi D_{\bar\mu}(x)
.
\eeq
The last term in the above expression for $\langle \Phi (y_1)\Phi (y_2)
\rangle^{\rm c}_\lambda $, which we singled out by its own number (20), 
will be significant later on.

The Fourier transform of the two-point function reads
\bdi
\wt{\langle \Phi\Phi\rangle}^{\rm c}_\lambda (p)= \frac{1}{p^2 +\bar\mu^2} -
\frac{2\pi \lambda^2}{(p^2 +\bar\mu^2)^2} \Bigl[ \cos 2\theta
(E_+ (\bar\mu) +\wt E^{(1)}_+ (p,\bar\mu))
\edi
\beq
+E_- (\bar\mu) - \wt E^{(1)}_- (p,\bar\mu) -8\pi^2 (\cos 2\theta +1)
\wt{D^2_{\bar\mu}}(0)\Bigr]
\eeq
\beq
\simeq \frac{1}{p^2 +\bar\mu^2 +2\pi\lambda^2 (\cos 2\theta (E_+ (\bar\mu)
+\wt E^{(1)}_+ (p,\bar\mu)) +E_- (\bar\mu) - \wt E^{(1)}_- (p,\bar\mu)
-\frac{2\pi}{\bar\mu^2}(\cos 2\theta +1)} ,
\eeq
where we inverted the two-point function perturbatively and used
\beq
\wt{D^2_{\bar\mu}}(0) =\frac{1}{4\pi \bar\mu^2}
\eeq
($\wt{D^2_{\bar\mu}}(p)$ is just a two-boson blob).

The Schwinger mass is given by the zero of the denominator in (23), which
leads to the equation (after rescaling, e.g., $\wt E^{(1)}_\pm (p,\bar\mu)
=\frac{1}{\bar\mu^2}\wt E^{(1)}_\pm (\frac{p}{\bar\mu},1)$, etc.)
\beq
-\frac{p^2}{\bar\mu^2} =1+\frac{2\pi \lambda^2}{\bar\mu^4} \Bigl[ \cos 2\theta
(E_+ (1) +\wt E^{(1)}_+ (\frac{p}{\bar\mu},1)) + E_- (1)-\wt E^{(1)}_-
(\frac{p}{\bar\mu},1) -2\pi (\cos 2\theta +1)\Bigr]
\eeq
with the second order solution ($M_2$ $\ldots$ second order Schwinger mass)
\bdi
M^2_2 :=-p^2 |_{\lambda^2} =\bar\mu^2 +\frac{2\pi \lambda^2}{\bar\mu^2}
\Bigl[ \cos 2\theta (E_+ (1) +\wt E^{(1)}_+ (i,1))
\edi
\bdi
+E_- (1) -\wt E^{(1)}_- (i,1) -2\pi (\cos 2\theta +1)\Bigr]
\edi
\beq
= \bar\mu^2 +\frac{2\pi \lambda^2}{\bar\mu^2} (2\pi \cos 2\theta A_+ +2\pi
A_- -2\pi (\cos 2\theta +1)) ,
\eeq
where
\beq
A_\pm =\int_0^\infty dr r \Bigl[ e^{-2K_0 (r)} -1 +I_0 (r)(\pm e^{\mp 2K_0 (r)
}\mp 1+2K_0 (r))\Bigr]
\eeq
\beq
A_+ =-0.6599,\qquad A_- =1.7277
\eeq
see \cite{SMASS,MSMPT} (we changed notation from $A,B$ in \cite {SMASS,MSMPT}
to $A_+,A_-$ here). Re-expressed in $\mu ,m$ we find
\bdi
M^2_2 =\mu^2 +2e^{2\gamma}\cos^2 \theta \, m^2 +2e^\gamma \cos\theta \, m
\sqrt{\mu^2 +e^{2\gamma} \cos^2 \theta \, m^2}
\edi
\bdi
+ e^{2\gamma}m^2 (A_+ \cos 2\theta +A_-) -e^{2\gamma}m^2 (\cos 2\theta +1)
\edi
\beq
= \mu^2 +2e^\gamma \cos\theta \, m\sqrt{\mu^2 + e^{2\gamma}\cos^2 \theta 
\, m^2} +e^{2\gamma} m^2 (A_+ \cos 2\theta +A_-) .
\eeq
Observe that one piece of the second order (in $\lambda$) contribution, 
stemming from the term that we singled out in (20), has precisely cancelled
the $o(m^2)$ piece of the lowest order (in $\lambda$) contribution, 
$\bar\mu^2$. This cancellation ensures that $M^2$, when Taylor expanded in
$m$ up to second order, coincides with the second order result of mass
perturbation theory (MPT),
\beq
(M^{\rm MPT}_2)^2 =\mu^2 +2e^\gamma \cos\theta \, \mu m 
+e^{2\gamma} m^2 (A_+ \cos 2\theta +A_-),
\eeq
see \cite{SMASS,MSMPT}.

Numerically, when evaluated at $\theta =0$ and for large $m$, 
\beq
\lim_{m\to\infty} M_2 (m,\theta =0) =me^\gamma (2+A_+ +A_-)^{1/2}
\simeq 3.12 m,
\eeq
we find that $M_2 (m=\infty ,\theta =0)$ is still about 55\% off the true
value. Therefore, a higher order computation would be highly desirable.
However, already the third order computation is very complicated. Still,
we are able to compute a part of the third order contribution with
considerably less effort. To achieve this aim, we use the requirement 
that the perturbation series in $\lambda$, when Taylor expanded in $m$, 
has to coincide with the mass perturbation series up to the given order.
The lowest order contribution (in $\lambda$) to $M^2$, $\bar\mu^2$, 
contains a third order (in $m$) piece,
\beq
\bar\mu^2 |_{m^3} =e^{3\gamma}\cos^3 \theta \, \frac{m^3}{\mu} ,
\eeq
and such a term cannot be produced by ordinary mass perturbation theory.
Therefore, the third order (in $\lambda$) contribution must contain a term
that, when Taylor expanded in $m$, precisely cancels the term (32). The
contribution that does the job reads
\beq
\delta_1 M^2_3 = -(2\pi)^3 \cos^3 \theta \, \frac{\lambda^3}{\bar\mu^4}=
-e^{3\gamma}\cos^3 \theta \, \frac{m^3}{\bar\mu} .
\eeq
A further contribution of order $\lambda^3$ may be identified by the same
argument via a closer inspection of ordinary mass perturbation theory in
order $m^3$. For this, observe that one contribution to our second order
result (26) stemmed from the evaluation of $\wt E_\pm^{(1)}(\frac{p}{\bar\mu},
1)$ at $(p/\bar\mu)^2 =-1$. For second order mass perturbation theory an
identical term occurs, but this time evaluated at $(p/\mu)^2 =-1$.
In addition, in mass perturbation theory there is a first order contribution
$\delta (M^{\rm MPT}_1)^2 =2e^\gamma \cos\theta\mu m$, therefore there must be
a third order contribution (in mass perturbation theory!)
\beq
\delta (M^{\rm MPT}_1)^2 \frac{\partial}{\partial (-p^2)}
(\cos 2\theta \, \wt E_+^{(1)}(p,1) -\wt E_-^{(1)}(p,1))|_{p^2 =-1}
\eeq
or, including all coefficients,
\bdi
-(2e^\gamma \cos\theta \mu m)\frac{e^{2\gamma}m^2}{2\pi \mu^2}\frac{\partial
}{\partial p^2}(\cos 2\theta \, \wt E_+^{(1)}(p,1) -
\wt E_-^{(1)}(p,1))|_{p^2 =-1}
\edi
\bdi
=-\frac{e^{3\gamma}\cos\theta \, m^3}{\pi \mu}\frac{\partial}{\partial p^2}
\int dx e^{ipx}(\cos 2\theta \, E_+^{(1)} (x,1) -E_-^{(1)}(x,1))|_{p^2 =-1}
\edi
\beq
=2e^{3\gamma}\cos\theta \, \frac{m^3}{\mu}(\cos 2\theta B_+ -B_-)
\eeq
\beq
B_\pm =\int drr^3 I_0 (r)E_\pm^{(1)}(r,1)
\eeq
\beq
B_+ =1.449,\qquad B_- =1.912 .
\eeq
This third order contribution must be matched by a corresponding third
order term in $\lambda$ perturbation expansion (observe that this term in
$\lambda$ perturbation theory cannot be produced in the same way, because
there is no first order correction $\delta M_1$ in $\lambda$ perturbation
theory; most likely, the term is produced by the $\langle \Phi\Phi \Phi^2
\exp\Phi\exp\Phi \rangle$ contributions). This corresponding term reads
\beq
\delta_2 M^2_3 =2(2\pi)^3 \cos\theta \, \frac{\lambda^3}{\bar\mu^4} 
(\cos 2\theta \, B_+ -B_-) =2\cos\theta \, 
e^{3\gamma}\frac{m^3}{\bar\mu}(\cos 2\theta \, B_+ -B_-) .
\eeq
Therefore, our new, partially third order corrected expression for the
Schwinger mass is 
\beq
M^2_{2\frac{1}{2}} =M^2_2 +\delta_1 M^2_3 +\delta_2 M^2_3 .
\eeq
Numerically, for $\theta =0$, $M_{2\frac{1}{2}}$ has the large $m$ limit
\beq
\lim_{m\to\infty} M_{2\frac{1}{2}}(m,\theta =0) =2.58 m ,
\eeq
which is nearly 30\% above the exact value $2m$. 

Of course, there are still some third order contributions missing, but as 
they are due to higher-dimensional convolution integrals of exponentially
decaying functions, we may hope that their contribution is not too large. 

We display the results of our renormal-ordered chiral perturbation theory
(for $\theta =0$)
in Fig. 2, and compare them with the lattice data of \cite{CH1,CK1} in
Fig. 3. Fig. 3 shows that our result $M_{2\frac{1}{2}}$ agrees with the 
lattice data very well up to $m\sim 1$; for larger $m$ the deviation increases,
reaching the above-mentioned 30\% for $m\to \infty$.

For an overview on additional  
data for the Schwinger mass (e.g., from light-front computations)
we refer to \cite{HHS} (especially their Fig. 2 and Table 16), and to
\cite{VFP}.

\section{Summary}

By a simple renormal-ordering of the bosonic Lagrangian of the massive
Schwinger model we were able to find a new perturbation expansion parameter,
$\frac{\lambda}{\bar\mu^2}$ (see (10), Fig. 1), that remains rather small
even for large fermion mass $m$. The resulting chiral perturbation theory
is similar to mass perturbation theory \cite{SMASS,MSMPT}, although it involves
somewhat more tedious computations. We used this renormal-ordered chiral 
perturbation theory to compute the Schwinger mass $M$, including all second
order (in $\frac{\lambda}{\bar\mu^2}$) and some third order contributions.
The resulting expression for the Schwinger mass $M$ reduces to the result
from mass perturbation theory (\cite{SMASS}) when Taylor-expanded in the
fermion mass $m$. For large $m$ our result describes the known data for the
Schwinger mass (e.g. the lattice data from \cite{CH1}) qualitatively
correct, although there remains some  deviation (e.g. our result $M_{2
\frac{1}{2}}$ differs by nearly 30\% from the known value $2m$ at the
far end $m\to \infty$).

Actually, we may conclude from Fig. 3 that high order (in $\frac{\lambda}{
\bar\mu^2}$) perturbative contributions will be relevant for $m\to\infty$.
The reason for this is that the expansion parameter $\frac{\lambda}{\bar\mu^2}
$ is nearly constant for $m\in [1,\infty ]$ , see Fig. 1. Therefore, low order 
contributions to $M$ will shift the curves for $M_{2\frac{1}{2}}$ (or $M_2$)
in Fig. 3 by a nearly constant amount for $m\in [1,\infty ]$. However, the
curves in Fig. 3 are closer to the lattice data at $m=1$ than at $m=\infty$,
and they can only approach the lattice data everywhere when high orders
in $\frac{\lambda}{\bar\mu^2}$ give a non-negligible contribution.

Finally, it should be mentioned that it could be interesting to apply this
renormal-ordered chiral perturbation theory to the computation of other
physical quantities of the massive Schwinger model (e.g., higher bound
states, chiral condensate). In addition, it might be possible to use a
version of this renormal-ordered chiral perturbation theory for the
massive multi-flavour Schwinger model, where ordinary mass perturbation
theory fails due to infrared divergencies \cite{Gatt1,Gatt2}.

\section*{Acknowledgement}

The author thanks the members of the Department of Mathematics at Trinity
College, where this work was performed, for their hospitality. Further
thanks are due to T. Heinzl and C. Stern for the data of \cite{CH1} and
\cite{CK1} in Fig. 3.

\begin{figure}
$$\psbox{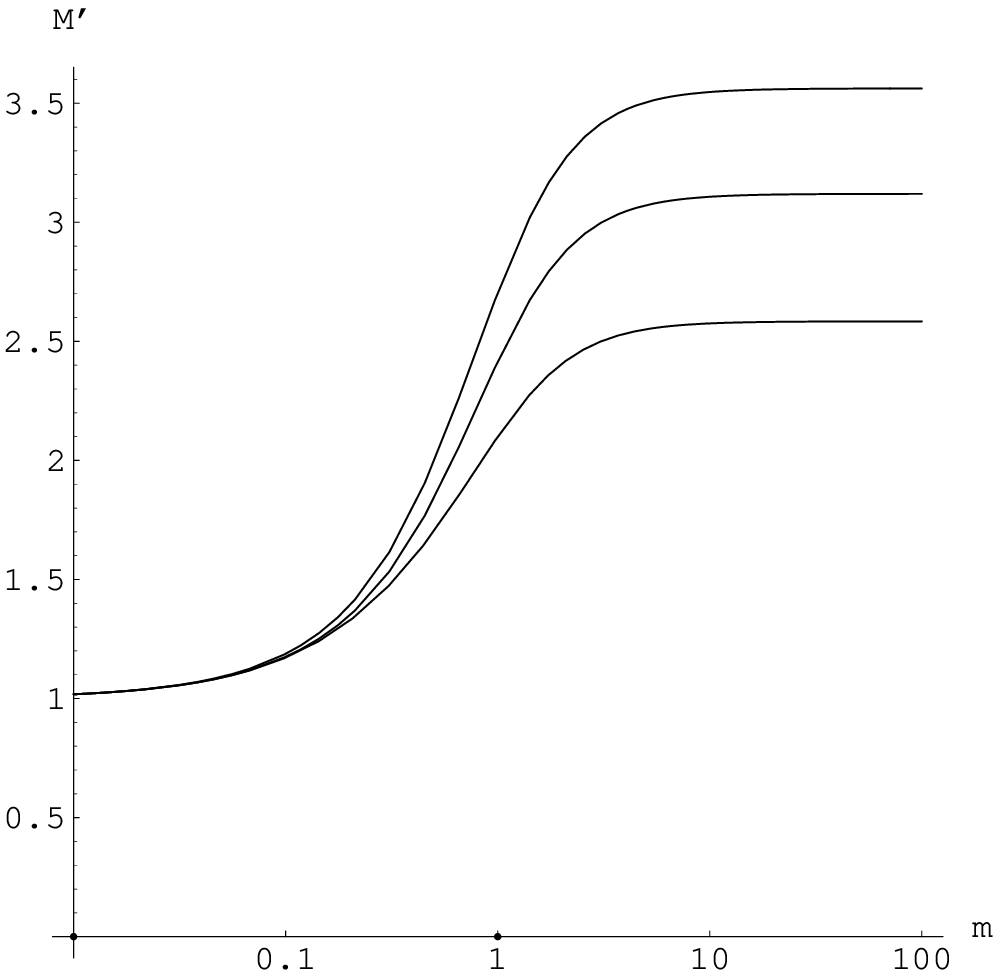}$$
\caption{The figure shows successive perturbative results for the Schwinger
mass. $\bar\mu$ is the upper curve, $M_2$ is the curve in the middle and
$M_{2\frac{1}{2}}$ is the lowest curve. We use the rescaled Schwinger mass
$M' =M/(1+m^2)^{1/2}$, like in \cite{HHS,VFP}, and choose units $\mu =1$.} 
\end{figure}

\begin{figure}
$$\psbox{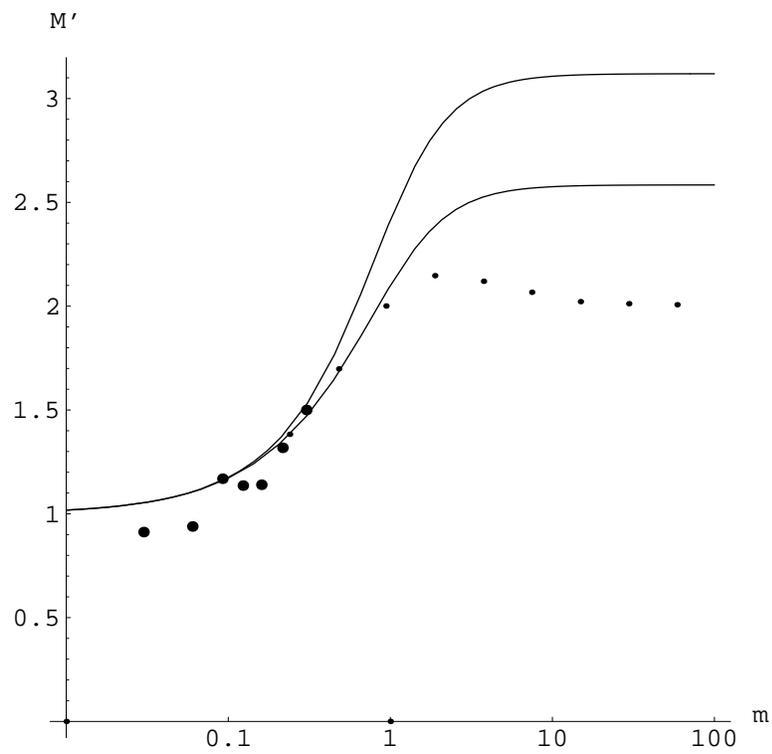}$$
\caption{The figure compares our results $M_2$, $M_{2\frac{1}{2}}$ to the 
lattice data of \cite{CH1} (small dots, for larger $m$) and \cite{CK1}
(fat dots, for small $m$, with rather large errors). We again use the 
rescaled Schwinger mass $M' =M/(1+m^2)^{1/2}$ and units $\mu =1$.}
\end{figure}

\end{document}